\documentclass[prb,twocolumn,showpacs,amsmath,amssymb]{revtex4}
\usepackage{color}
\usepackage{graphicx}
\usepackage{dcolumn}
\usepackage{bm}

\begin{document}

\title{Vortex matter in mesoscopic two-gap superconducting disks:\\ influence of Josephson and magnetic coupling}

\author{R. Geurts}
\email{roeland.geurts@ua.ac.be}
\author{M. V. Milo\v{s}evi\'{c}}
\email{milorad.milosevic@ua.ac.be}
\author{F. M. Peeters}
\email{francois.peeters@ua.ac.be}

\affiliation{Departement Fysica, Universiteit Antwerpen,
Groenenborgerlaan 171, B-2020 Antwerpen, Belgium}

\date{\today}

\begin{abstract}
The effects of the coupling between two electronic condensates in
two-gap mesoscopic superconductors are studied within the
Ginzburg-Landau theory using a finite difference technique. In
applied magnetic field, we derive the dependency of the size of the
vortex on the sample size and the strength of the Josephson
coupling. In addition, we elaborate on the dependence of the
critical temperature and field on the parameters of the coupled
condensates. We demonstrate further the existence and stability of
fractional states, for which the two condensates comprise different
vorticity. Moreover, we also found pronounced asymmetric fractional
states and we show their experimentally observable magnetic
response. Finally we introduce the magnetic coupling between
condensates, and study in particular the case where one band is type
II and the other is type I, i.e. the sample is effectively of I.x
type. The calculated $M(H)$ loops show a clear signature of the
mixed type of superconductivity, which we find to be strongly
affected by the ratio of the coherence lengths in the two
condensates.
\end{abstract}

\pacs{74.78.Na, 74.25.Dw, 74.25.Qt}

\maketitle

\section{Introduction}
MgB$_2$ is the first superconductor unambiguously shown to possess
two superconducting gaps.\cite{MgB2TwoGap} Since its discovery in
2001,\cite{MgB2Discovery} a lot of research was conducted on this
specific material as well as on two-band superconductors in general.
In its class of binary compounds and metallic superconductors,
MgB$_2$ turns out to have the highest critical temperature known
today, $T_c=39$ K. Its bulk critical field is strongly anisotropic:
$3.5$ T along the $c$-axis of the crystal and $17$ T in the
$ab$-plane and can reach as high as $43$ T in
films.\cite{Hc2Anisotropy,MgB2SCReview, MgB2SCReview2}

While the mechanism of its superconductivity is not yet entirely
understood, it has been experimentally proven that MgB$_2$ has two
separate superconducting gaps. For example, in Refs.
\cite{PiBandVortexImaging, STMPerBand} the separately imaging of the
$\pi$- or $\sigma$-bands was demonstrated. On the theoretical side,
one considered two order parameters to describe the superconducting
properties of MgB$_2$. One of the first Ginzburg-Landau (GL)
descriptions of multigap superconductors was developed by
Zhitomirsky and Dao, starting from microscopic theory.\cite{Drag5}
Fitting to experimental results, the authors pinpointed the values
of several GL parameters relevant for MgB$_2$, and derived
analytical expressions for the critical parameters. In Ref.
\cite{Hc2AnisotropyGL} the same authors discussed the anisotropy of
$H_{c2}$ within the GL-framework. They considered only the direct
exchange of Cooper pairs between condensates, i.e. the so called
`Josephson' coupling. In the work of Askerzade {\it et al.}
 a different interaction between the bands was
investigated - the drag effect,\cite{Drag3,Drag4} which is described
in the GL-formalism through the coupling of the gradient terms of
the two condensates. The apparent agreement with experiment led
these authors to fitting parameters for the GL-model of MgB$_2$.

MgB$_2$ is generally accepted to be a type-II superconductor.
However, in a very clean sample, Moshchalkov {\it et al.} estimated
that one of the bands could be type-I and the other
type-II.\cite{Mosh,Squid} The resulting system exhibits behavior
that cannot be attributed to either type, thus the name type 1.5
superconductivity seemed credible. Indeed, the authors found a
strong clustering of vortices, a phenomenon which they ascribe to a
{\it combination} of attractive and repulsive vortex-vortex
interaction. Actually, Ref. \cite{TypeIXSurfaceEnergy} reported a
positive surface energy for vortices whenever the coherence lengths
of the two bands are comparable. In Ref. \cite{SemiMeissner} the
semi-Meissner state was predicted theoretically for a two-gap
superconductor and for superconductors which do not belong to either
of the two classes type-I or type-II. The possibility of vortices
carrying non-integer flux was studied in Refs.
\cite{FractionalFlux}.

Surprisingly, virtually all studies done on two-gap superconductors
(TGS) up-to-date concern bulk samples. It is known however that
mesoscopic superconductivity bears a number of fascinating
phenomena, ranging from specific vortex states, to enhancement of
critical parameters by quantum tailoring. The only existing example
of such a study is the work of Chibotaru {\it et al.} on mesoscopic
disks.\cite{Chib2Gap} As a novelty, the authors found that
fractional vortex states (when bands have different vorticity) can
be realized in a TGS and can even be thermodynamically stable.
However, those results turned out to be specific to the case of very
weak coupling and not realistic for MgB$_2$.

In this paper we analyze the fundamental properties and vortex
matter of mesoscopic disk-shaped two-gap superconductors using the
Ginzburg-Landau formalism, where the electronic exchange between
condensates occurs through Josephson coupling, and magnetic exchange
between condensates is allowed for. The latter mechanism has not yet
been studied in detail up to now. The paper is organized as follows.
In Section II, after describing the theoretical approach, we focus
on the effects of Josephson coupling on the the size of a vortex
core, the unique vortex states and their $H-T$ stability regions,
and the critical temperature and field as a function of coupling
strength. In Section III, we introduce the screening of the magnetic
field into the theoretical formalism, and illustrate the influence
of the magnetic coupling between the two condensates on the vortex
states, particularly in the case of type I.x superconductivity.
Magnetic signatures of the different features are discussed in the
light of potential observation by magnetometry. Finally, our
findings are summarized in Sec. IV.

\section{Josephson coupling}

\subsection{Theoretical formalism}

It is widely accepted that the high critical temperature of MgB$_2$
arises due to the coupling of the superconducting bands which
effectively reinforce each other. However, the exact nature of the
coupling is not fully understood, and possible scenarios are the
exchange of electrons, Cooper pairs, interaction between the
respective supercurrents, interaction through the internal magnetic
field, etc. Microscopic ab-initio calculations have not been able to
pinpoint the key interaction. In what follows, we will consider the
Josephson coupling between the bands, resulting from the tunneling
of the Cooper pairs from one band to another. This is incorporated
in the Ginzburg-Landau (GL) energy functional \cite{Drag5} through
an interaction term dependent on the order parameter of both bands
and proportional to $\Gamma$, the Josephson coupling strength:
\begin{widetext}
\begin{equation}
 \Delta F = \int \left[ \sum_{n=1}^{2} \left( \frac{1}{2m_n} \left|
(-i\hbar\nabla - \frac{2e}{c} \vec{A}) \Psi_n \right|^2 + \alpha_n
|\Psi_n|^2 + \frac{1}{2} \beta_n |\Psi_n|^4 \right) - \Gamma
(\Psi_{1}^{*}\Psi_2 + \Psi_{2}^{*}\Psi_1) ) \right] dV,
\label{GLFunc}
\end{equation}
\end{widetext}
where $\alpha_n = \alpha_{n0} (1-T/T_{cn})$ and $\beta_n$ are the GL
coefficients, and $n$ is the band index. This results in a set of
nine parameters describing a two-gap system: $\alpha_{10}$,
$\alpha_{20}$, $\beta_1$, $\beta_2$, the Cooper-pair mass $m_1$ and
$m_2$, $\Gamma$, and the critical temperatures $T_{c1}$ and
$T_{c2}$. While each band has its own intrinsic critical
temperature, Josephson coupling causes both bands to survive up to a
higher critical temperature, $T_c > \max(T_{c1},T_{c2})$. While the
GL functional is derived for $T \lesssim T_c$, experience with
mesoscopic single-gap superconductors indicates that the GL
equations in practice are valid much deeper into the superconducting
state. In Eq. (\ref{GLFunc}) we then introduce temperature
independent units, in order to rewrite it in a dimensionless form.
We express the free energy of the system in units of
$F_{10}=\alpha_{10}^{2}/\beta_1$, length in units of $\xi_{10}$
($\xi_{n0}=\hbar/\sqrt{-2m_n\alpha_{n0}}$), the vector potential in
$A_0=\hbar c / 2 e \xi_{10}$, the order parameters in
$\Psi_{n0}=\Psi_{n0}(T=0, \Gamma=0,
H=0)=\sqrt{-\alpha_{n0}/\beta_n}$ and the temperature in $T_{c1}$.
This reduces the set of necessary parameters to five:
$\delta=\Psi_{10}/\Psi_{20}$, $\alpha=\xi_{10}^{2}/\xi_{20}^{2}$,
$m=m_1/m_2$, $T_{cr}=T_{c2}/T_{c1}$ and $\gamma =
\Gamma/\alpha_{10}$. In addition, we have two external tuneable
parameters: the temperature $T$ and the applied field $H$. The flux
$\phi$ is defined as the externally applied flux. We first consider
an extreme type-II case, and neglect the self-induced magnetic field
in the sample.

The minimization of the energy functional leads to the two-band GL
equations. After the scaling described above, the equations for the
order parameters read:
\begin{equation*}
\left\{ \begin{aligned} (-i\nabla - \vec{A})^2 \psi_1 - (1 - T -
|\psi_1|^2) \psi_1 -
\frac{\gamma}{\delta} \psi_2 = 0, \\
\frac{1}{\alpha}(-i\nabla - \vec{A})^2 \psi_2 -
\left(1-\frac{T}{T_{cr}} - |\psi_2|^2 \right) \psi_2  -
\frac{\gamma\delta}{m\alpha}\psi_1 = 0.
\end{aligned}
\right. \label{glequat}
\end{equation*}
This system of non-linear coupled differential equations we solve
numerically on a square grid of typically 128x128 points. The
details of this procedure can be found in Ref. \cite{GLNumerical}.

In the following analysis we neglect the screening of the magnetic
field. This is justified for an extreme type-II material, or any
sufficiently thin sample. An applied vector potential
$A=(\frac{1}{2}Hy,-\frac{1}{2}Hx,0)$ results in a magnetic response
resulting from a total supercurrent:
\begin{equation}
\vec{j}_s =\Re\left[ \psi_1\left( i\nabla -
\vec{A}\right)\psi_{1}^{*} \right] + \frac{m}{\delta^2} \Re\left[
\psi_2\left( i\nabla - \vec{A}\right)\psi_{2}^{*} \right].
\label{SuperCurrent}
\end{equation}
The free energy functional in dimensionless units reads:
\begin{eqnarray}
\frac{\Delta F}{\alpha_{10}^{2}/\beta_1} &= \int&  \Big[ \left|
(-i\nabla - \vec{A}) \psi_1 \right|^2
- (1-T) |\psi_1|^2 + \frac{1}{2} |\psi_1|^4 \nonumber \\
& &  + \frac{m}{\delta^2}  \left| (-i\nabla - \vec{A})
\psi_2 \right|^2   \nonumber \\ & &+ \frac{m\alpha}{\delta^2} \left( -(1-T/T_{cr}) |\psi_2|^2  + \frac{1}{2} |\psi_2|^4 \right)  \nonumber \\
& & - \frac{\gamma}{\delta} (\psi_{1}^{*}\psi_2 +
\psi_{2}^{*}\psi_1) )  \Big] dV .\label{FGL}
\end{eqnarray}

Let us here address several direct implications of Josephson
coupling. It is clear from Eq. (\ref{GLFunc}) that the sign of
$\gamma$ determines the relative phase shift between the order
parameters in the two condensates - either $\approx 0$ when
$\gamma>0$ or $\approx \pi$ when $\gamma<0$ - in order for the
coupling term to provide a negative energy contribution. However,
the sign of $\gamma$ has no influence on observables such as the
Cooper-pair density and magnetic response of the sample. The general
consequence of $\gamma$ coupling is an injection of Cooper pairs
from one band into the other and vice versa, thus increasing the
stability of the superconducting state. In other words, the average
Cooper-pair density always increases with $\gamma$. In the absence
of an applied field, the ratio $\chi=\psi_1/\psi_2$ can be found
from
\begin{equation}
\frac{\gamma \delta}{m \alpha} \chi^4 +
\left(1-\frac{T}{T_{cr}}\right)\chi^3 - \left(1-T\right)\chi -
\frac{\gamma}{\delta}=0, \label{eqpsiratio}
\end{equation}
analytically derived from the GL-equations. From this we find that
in the limit $\gamma \rightarrow \infty$ a constant ratio
$\psi_{1}/\psi_{2}($H=0$)=\sqrt{\sqrt{m\alpha}/\delta}$ is reached,
independent of temperature.

\begin{figure}[t]
\includegraphics[width=\linewidth]{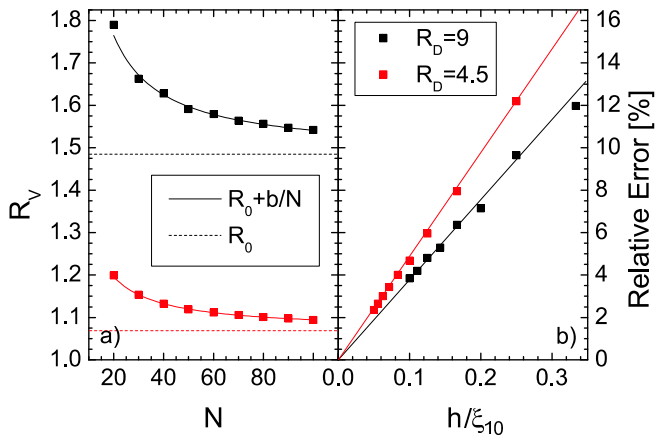}
\caption{\label{VortexSizeGridResolution} (Color online) (a) The
dependence of the observed vortex size $R_V$ on the number of grid
points in the numerical mesh $N$. (b) The relative error in found
vortex size vs. grid spacing $h$, for two different sizes of the
superconducting disk. }
\end{figure}

The next section is dedicated to an analysis of the size of the
vortex core. Before we get into the physics of the problem, we here
address some numerical issues following from mapping of the
superconducting {\it disk} on a {\it square} numerical grid. For
obvious reasons, the influence of the resolution of the numerical
grid $N$, on the observed vortex size $R_V$, is significant. In Fig.
\ref{VortexSizeGridResolution}(a) we show the $R_{V}$ vs. $N$
(definition of $R_V$ is given in the next section). With increasing
grid density, the numerical error decreases, and the vortex size
converges towards the $R_0$ value with dependence $R_V = R_0 + b/N$.
By a fitting procedure $R_0$ and $b$ can be determined and the
corresponding curves are represented by the solid lines in the
figure. The dashed lines are the asymptotes with value $R_0$. In
Fig. \ref{VortexSizeGridResolution}(b) the relative error is plotted
as a function of $h$, the grid spacing (proportional to sample size
and inversely proportional to $N$). We conclude that due to numerics
the vortex size is always slightly overestimated, with overshoot
increasing for smaller samples (where influence of the boundaries is
more pronounced). Nevertheless, with a resolution of 10 points per
coherence length we get a relative error under 5\%. Although higher
grid density obviously improves the results, we refrain from using a
density above 10 points/$\xi$, in order to optimize the speed of the
calculation.

\subsection{Size of the vortex core}

The coherence length is the characteristic length scale over which
the order parameter changes. It is therefore intuitive that the size
of the vortex core is proportional to the coherence length in bulk
superconductors. In single-gap materials, the coherence length is
proportional to $1/\sqrt{1-T/T_c}$ in the temperature range where
the Ginzburg-Landau theory is valid. Here we show that in two-gap
superconductors the coherence length is strongly affected by the
coupling parameter $\gamma$, generally in an opposite manner from
temperature. For comparison, we can use the coherence length
obtained from the expression for the second critical field derived
in Ref. \cite{Drag5}, through the relation
$H_{c2}=\Phi_0/2\pi\xi^2$, with $\Phi_0$ being the flux quantum.
Deviations are a priori expected, since already experiments of Refs.
\cite{PiBandVortexImaging, STMPerBand} found a discrepancy between
the vortex size and the coherence length deduced from the second
critical field.

To estimate the coherence length, we will numerically determine the
size of the vortex core, a quantity which is not uniquely defined.
In the following calculations, we examine the single-vortex state in
a MgB$_2$ superconducting disk exposed to a field providing three
flux quanta through the sample. For the definition of the vortex
size, two possibilities are considered in the literature: (i) The
vortex size is determined by the distance from the center of the
vortex to the contour where the Cooper-pair density (CPD) recovers
to some percentage of its maximal value in the sample, denoted as
$R_{V,CPD}$; and (ii) The vortex size is the distance from the
center of the vortex to the first contour where the supercurrent
$j_s$ reaches its maximum, denoted as $R_{V,j}$. The problem of the
first definition is the arbitrary threshold value for the criterion,
but also the fact that we have two Cooper pair densities, which
makes the single vortex size ambiguous. As a threshold we take 80
\%, since this allows for a more precise vortex size determination,
and we will consider only the first condensate. On the other hand,
the second definition involves coupled condensates and thus provides
us with a unique vortex size. We therefore adopt the second
definition to describe the vortex size in the rest of this work.
Contrary to the bulk case, in our mesoscopic disks both definitions
render a vortex size dependent on the radius of the disk $R_D$, as
vortex currents in the center of the sample can interact with
Meissner currents decaying from the edge inwards. However, while
$R_{V,CPD}$ in each band saturates for $R_D \rightarrow \infty$,
this is not the case for $R_{V,j}$: it develops a linear dependence
on $R_D$. We extracted the exact dependence of both definitions of
the vortex size $R_V$ on the disk size $R_D$, which led us to a
universal formula (valid for both definitions, but with different
coefficients):
\begin{equation}
\left( \frac{1}{R_{V0}} \right)^2 + \left( \frac{c}{R_D} \right)^2 =
\left( \frac{1}{R_V-hR_D}\right)^2, \label{VortexSize}
\end{equation}
\begin{figure}[b]
\includegraphics[width=0.8\linewidth]{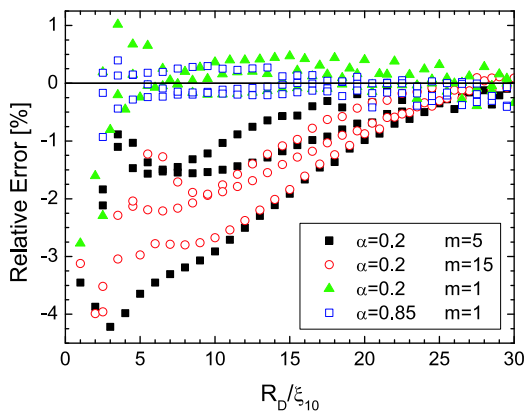}
\caption{\label{RVvsRDrelativeerror} (Color online) The relative
error of the result of Eq. (\ref{VortexSize}) compared to the
observed vortex size, as a function of disk size, for indicated
different values of parameters $m$ and $\alpha$, each with three
different combinations of $\gamma$ and $T$ in order to cover as
large as possible parameter space in the analysis.}
\end{figure}
\begin{figure}[b]
\includegraphics[width=\linewidth]{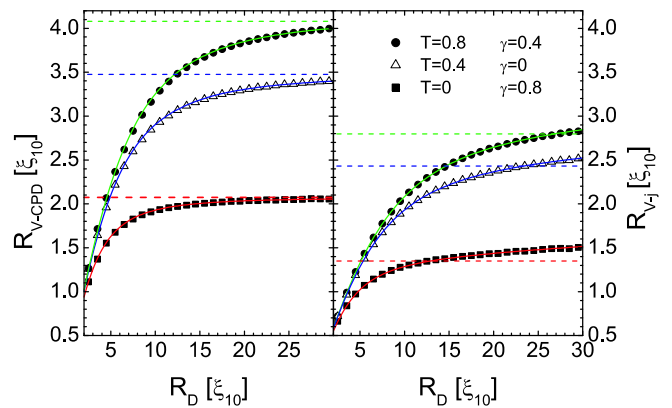}
\caption{\label{RVCPDVsRD2} (Color online) Vortex size, determined
through the decay of the Cooper-pair density $R_{V-CPD}$ (a) and
determined through the maximum of encircling currents $R_{V-j}$ (b),
as a function of the radius of the superconducting disk $R_D$. The
solid dots represent the numerical results, while solid lines show
the fit using Eq. \ref{VortexSize}. The dashed lines indicate
$R_{V0}$, the fitting parameter corresponding to the
sample-independent vortex size.}
\end{figure}
where $R_{V0}$ is the vortex size independent of the sample size,
$c$ is a length coefficient and $h$ is the slope of $R_{V}$ vs.
$R_D$ for large $R_D$. From fitting of our numerical data, collected
at different $T$, $\gamma$, $\alpha$ and $m$, we obtained
\{$c=1.90$, $h=0$\} for $R_{V, CPD}$ and \{$c=3.40$, $h=0.006$\} for
$R_{V, j}$. These coefficients are valid for single gap
superconductors {\it as well as} for two-gap superconductors, even
when the coherence lengths of the two condensates are very different
(e.g. for small $\alpha$, see Fig. \ref{RVvsRDrelativeerror}). The
function gives an excellent estimate of the vortex size for
$m \neq 1$, for disks larger than $5\xi$. As
shown in Fig. \ref{RVvsRDrelativeerror}, deviation from the given
function does occurs for specific choices of $m$ and $\alpha$,
especially for small disks, but the relative error remains under
5\%.

Both previously given definitions of $R_V$ are illustrated in Fig.
\ref{RVCPDVsRD2}. When disks are too small, the vortex size is not
always fixed by the disk size only, e.g. when $T > T_{cr}$ and
$\gamma < 0.1$, when the coherence lengths differ much and the
interaction of the vortex with the Meissner currents becomes too
different in the two bands. We found however that when $R_D > 10
\xi_{10}$ these mesoscopic effects have only a minor influence and
the correspondence between $R_V$ and $R_D$ becomes predictable
again.

With the established dependence of the vortex size on the size of
the sample, we can more precisely determine the actual influence of
the coupling $\gamma$ on the vortex size. In particular we will look
at the behavior of $R_{V0, j}$ as a function of $T$, $T_{cr}$ and
$\gamma$ (parameters $\alpha$ and $m$ remain fixed at realistic
values for MgB$_2$).

\begin{figure}[t]
\includegraphics[width=\linewidth]{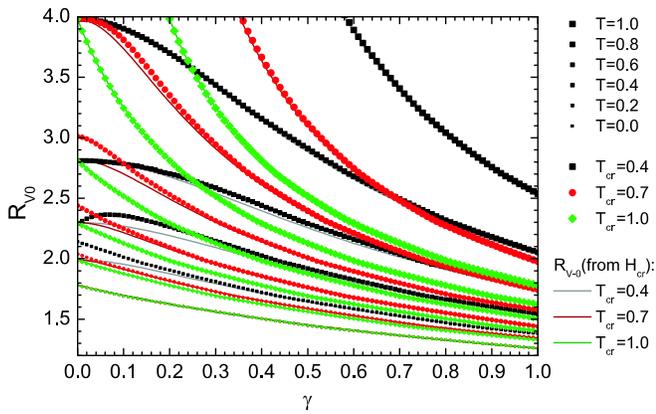}
\caption{\label{RV0vsGamma} (Color online) $R_{V0}$ as a function of
$\gamma$ and temperature, for $\gamma<1$ and for different values of
$T_{c2}$. $\alpha=m=1$.}
\end{figure}

In Fig. \ref{RV0vsGamma} the size-independent vortex size $R_{V0}$
is plotted versus $\gamma$, for different temperatures $T$ and
$T_{cr}$. The dots represent the result of our simulations. The
general behavior can be described by the following observed trends:
i) Increasing $T$ induces an increase of the vortex size whereas
increasing $\gamma$ has the opposite effect. ii) Deviation from the
latter monotonic behavior occurs when $T\approx T_{cr}$ and the
coupling is weak, see e.g. the curve at $T=T_{cr}=0.4$. The reason
for the initial positive slope is that the second band is revived by
the presence of coupling but retains its own character (i.e. a
larger coherence length) since coupling is still weak. iii) For $T
> T_{cr}$, curves with different $T_{cr}$ but identical $T$ merge at
$\gamma=0$ since then only the first condensate survives and fully
determines the vortex size. In Fig. \ref{RV0vsGamma}, the solid
curves represent an estimate of the vortex size based on the general
relation between the coherence length and the upper critical field
in a single gap bulk superconductor $H_{c2}\propto 1/\xi^2$,
\begin{equation}
\xi = \frac{\sqrt{2}}{\sqrt{ g_{+}(\alpha, T, T_{cr}) + \sqrt{
g_{-}(\alpha, T, T_{cr})^2 + 4 \frac{\gamma^2}{m}}}}, \label{xiHc2}
\end{equation}
with $g_{\pm}(\alpha, T, T_{cr})=1 - T \pm \alpha \left(
1-\frac{T}{T_{cr}}\right)$, based on the analytical expression for
the critical field of a bulk two-gap superconductor,
\begin{eqnarray}
H_{c2}(T) & \propto & 1 - T + \alpha \left(
1-\frac{T}{T_{c2}}\right) + \nonumber \\&& \sqrt{ \left( 1 - T -
\alpha \left( 1-\frac{T}{T_{c2}}\right) \right)^2 + 4
\frac{\gamma^2}{m}}, \label{hc2twogap}
\end{eqnarray}
taken from Ref. \cite{Drag5}. We find that the vortex size in our
samples scales with the coherence length as $R_{V0} = 1.78 \xi$,
which we use to plot the curves in Fig. \ref{RV0vsGamma}. These
theoretical curves coincide rather well with the data for $\gamma
>0.25$. The reason for this is that, when coupling becomes
sufficiently strong, both order parameters tend to have a similar
spatial distribution and thus also exhibit a similar vortex size and
coherence length. A good correspondence between the data and the
fitted curves is also found for $T \gg T_{cr}$, i.e. when the second
condensate exists solely due to the coupling to the first
condensate, or in the case of weak coupling and the second
condensate is almost depleted, so that it does not influence the
vortex size. Two regions of discrepancy include $T\approx T_{cr}$
($R_{V0}$ behaves non-monotonic), and $T \ll T_{cr}$ and weak
coupling. For the latter case, the formula still predicts
$R_{V0}(\gamma=0)$ to be independent of $T_{cr}$, while this is
clearly not the case. In this regime the vortex size is found to
behave more like that of a single gap superconductor, but with a
different critical temperature. By fitting we determined a function
that describes the behavior of the vortex size accurately in this
regime as
\begin{equation}
R_{V0}^{}=\frac{1.78}{\sqrt{1-T/\sqrt{T_{cr}}+\gamma}}.
\label{RV0Asymp}
\end{equation}
This equation is generally applicable, and effectively shows our
initial premise that $\gamma$ has an opposite influence to $T$.

\begin{figure}[b]
\includegraphics[width=0.8\linewidth]{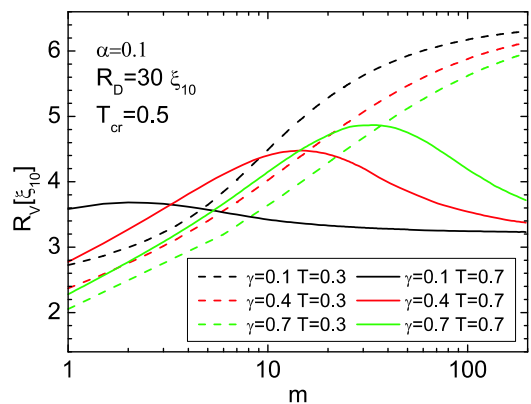}
\caption{\label{RVvsMassAlfa0.1} (Color online) The apparent vortex
size in a disk with radius $R=30\xi$ as a function of $m$, for
different $\gamma$ and $T$. The ratio of coherence lengths in two
condensates is fixed at $\xi_{20}/\xi_{10}=3.162$, i.e.
$\alpha=0.1$.}
\end{figure}

We notice however that Eq. (\ref{xiHc2}) also contains the
dependence on $\alpha$ and $m$. However, this formula can not
adequately describe the vortex size for $m,\alpha$ much different
from 1, since the properties of the two condensates can no longer be
described by a single coherence length. In general we can state that
the relation between the coherence length and the critical field
does not hold anymore when the individual coherence lengths differ
too strongly. In Fig. \ref{RVvsMassAlfa0.1} we plot the numerically
obtained $R_V$ as a function of $m$, for a small parameter $\alpha$
(with thus an acute difference between the coherence lengths in the
two condensates). The analytic estimate of Eq. (\ref{xiHc2}) is
monotonously increasing with $m$ in this case, and is obviously not
useful for comparison with non-monotonically evolving curves in Fig.
\ref{RVvsMassAlfa0.1}.

Let us first analyze the limiting case of extremely large $m$.
Following from Eq. (\ref{glequat}), the second condensate decouples
from the first one in this limit. At the same time, as seen in Eq.
(\ref{SuperCurrent}), the influence of the second condensate on the
total current in the system increases. Therefore, the size of the
vortex $R_{V,j}$ will be fully determined by the second condensate
(and its nominal coherence length $\xi_2(T)$), provided that the
temperature is below $T_{cr}$. Otherwise, the vortex size is
determined solely by the first condensate (and $\xi_1(T)$), since
the coupling between the condensates is entirely suppressed and the
second condensate fully depletes.

This helps us to understand the behavior of the vortex size as a
function of $m$, shown in Fig. \ref{RVvsMassAlfa0.1}. At $T<T_{cr}$,
the initial increase of $m$ decreases the coupling of the second
condensate to the first ($m$ has an opposite effect from $\gamma$,
see Eq. (\ref{glequat}). This causes an increase of the apparent
vortex size, due to much larger coherence length of the second
condensate ($\alpha=0.1$), and at the same time the increase of $m$
makes the supercurrent of the second band stronger and thus more
deterministic for the magnetically detectable size of the vortex
(see Eq. (\ref{SuperCurrent}). Above $T_{cr}$ these two effects
become competing, since the first will deplete the second condensate
and therefore reduce its influence while the second enhances the
influence of the second condensate. These competing effects result
in the non-monotonic behavior of the vortex size vs. $m$ in Fig.
\ref{RVvsMassAlfa0.1}. At low $m$, the influence of the large
coherence length in the second condensate dominates, whereas at
large $m$ the coupling disappears and the second condensate
depletes. At very large $m$, all curves for $T>T_{cr}$ saturate to
the same value, namely the size of the vortex core in the first
condensate, in the absence of a second one.

\begin{figure}[b]
\includegraphics[width=1\linewidth]{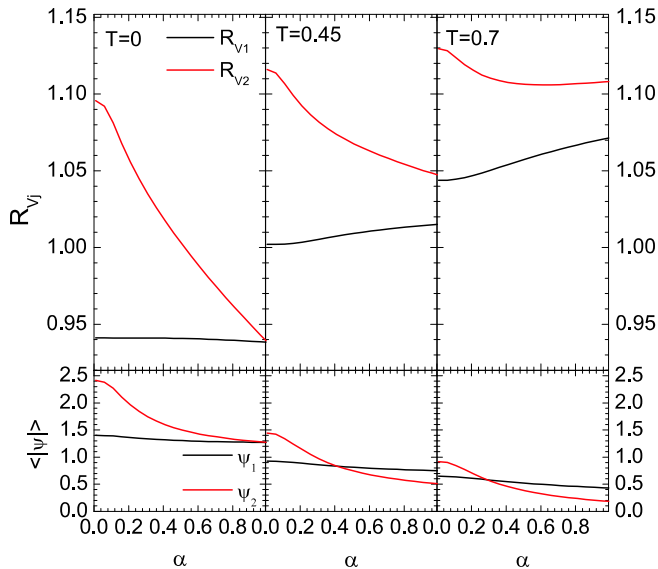}
\caption{\label{RVvsAlfa} (Color online) Top: $R_{V}$ in each
condensate as a function of $\alpha$, for different $\gamma$, $T$.
Bottom: Ratio of condensation energies of the two condensates
(logarithmic scale) vs. $\alpha$. When equal to $1$, condensates
influence each other with equal strength since $m=1$.}
\end{figure}

In Fig. \ref{RVvsAlfa} we demonstrate some peculiarities of the
dependence of the vortex size on the parameter $\alpha$. $\alpha$
was swept down from $1$ to $0.005$ in a disk with parameters
$m=\delta=1$, $T_{cr}=0.5$ and $R_{D}=4\xi_{10}$. In the absence of
coupling, this sweep increases $\xi_{20}$ while $\xi_{10}$ is kept
constant, as the length unit of the GL equations. In the presence of
coupling, both the resulting $\xi_2$ and the resulting $\xi_1$ will
be influenced (i.e. the Cooper-pair correlation length in each of
the condensates, different from the nominal coherence lengths in
each condensate separately). Intuitively, one expects that coupling
causes vortex cores in the two condensates to have similar behavior,
and tend towards similar sizes; instead, for decreasing $\alpha$ at
temperatures $T=0.45$ and $T=0.7$ an increase of $R_{V2}$ is
observed while $R_{V1}$ decreases! For stronger coupling this effect
becomes even more prominent, compared to the vortex size at
$\alpha=1$. At lower temperatures (shown for $T=0$ in Fig.
\ref{RVvsAlfa}), the behavior of $R_{V,j}(\alpha)$ for $T=0$ is in
better concordance with the intuition: $R_{V1}$ increases as
$R_{V2}$ increases, with the effect growing with coupling. However
this effect is weak.


To better understand the behavior of the vortex sizes we will invoke
the full free energy expression of Eq. (\ref{FGL}), which can be
written as $F=F_1+F_2+F_{12}$, where $F_i$ depends only on $\psi_i$
and $F_{12}$ is the Josephson coupling term. The key point here is
that a decrease of $\alpha$ will lower $F_2$ directly, therefore
giving more weigth to the other terms $F_1$ and $F_{12}$, i.e.
changing $\alpha$ reorders the hierarchy of the terms. A stronger
$F_{12}$ stimulates an increase of $\psi_1$ and $\psi_2$. Since
$F_2$, which regulates the size of $\psi_2$, becomes less important,
$\psi_2$ will increase much faster than $\psi_1$. A similar behavior
of the $\psi$ and $R_V$ curves hints to the link between the two
variables. For an infinite superconductor, sweeping $\alpha$ to zero
would cause both $\psi_1$ and $\psi_2$ to diverge. However for a
finite (mesoscopic) superconductor, the vortex size will eventually
exceed the disk size, thereby effectively suppressing the order
parameter and preventing the divergence.

The decrease of $R_{V1}$ with decreasing $\alpha$ can be ascribed to
the increase of $\psi_1$. However, in the left panel of Fig.
\ref{RVvsAlfa} where temperature is zero, we notice the subtle
increase of $R_{V1}$ with decreasing $\alpha$. Here $\psi_2$ is
still relatively large (because of the low temperature) which causes
$\psi_1$ to feel a strong influence from $\psi_2$ due to coupling.
As a result, for low $T$, vortex sizes will tend to be similar in
the two condensates, as we intuitively predicted. In the central and
the right panel of Fig. \ref{RVvsAlfa}, i.e. for higher
temperatures, the influence of $\psi_2$ is reduced, and the
increased order parameter $\psi_1$ prevails in determining the
vortex size - thus $R_{V1}$ decreases. In the right panel $T>T_2$,
and the second condensate would be depleted in the absence of
coupling. This creates yet another regime of the $R_V(\alpha)$
dependence: initially $R_{V2}$ now decreases with decreasing
$\alpha$ from 1. The second condensate is completely dependent on
the first one, and therefore obeys its shape. When $\alpha$ further
decreases, $\psi_2$ is less modified and an increase of $R_{V2}$, is
recovered.

\begin{figure}[h]
\includegraphics[width=\linewidth]{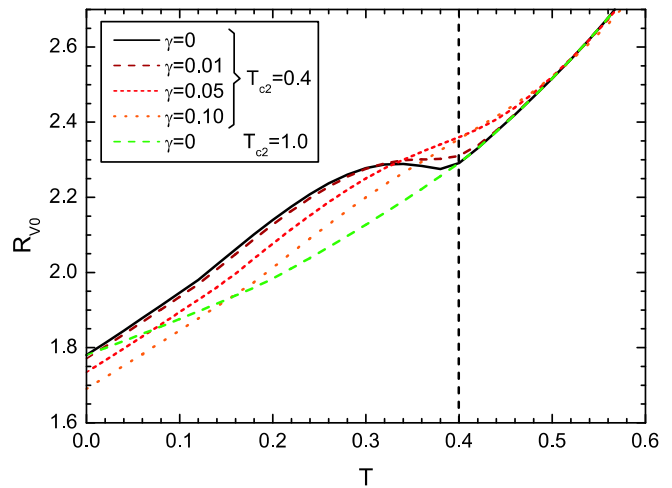}
\caption{\label{RinfVsTemperature} (Color online) The
size-independent vortex size $R_{V0}$ as a function of temperature
(for fixed $T_{c1}$ and $T_{c2}$), for several values of $\gamma$.
$R_D=10\xi_0$, $m=1$, and $\alpha=1$.}
\end{figure}

Finally we point out one more interesting artefact. In Fig.
\ref{RinfVsTemperature} we show the calculated vortex size as a
function of temperature, for very weak coupling, i.e. small
$\gamma$, and $m=\alpha=1$. The observed kink corresponds to the
critical temperature of the second band, and smears out when
$\gamma$ is increased. The \{$\gamma=0, T_{cr}=0.4$\} curve starts
at the same value as the \{$\gamma=0, T_{cr}=1$\} curve, since at
$T=0$ there is no dependence of vortex size on $T_{cr}$. This
behavior is observable by magnetic-force, scanning Hall probe, or
scanning tunneling microscopy, and we expect its experimental
verification.

For the \{$\gamma=0, T_{cr}=1$\} curve, the superconducting state
and supercurrents in both condensates are identical. Therefore the
vortex size defined on separate condensates as well as on the
combined system will be equal. For this reason the curve coincides
with the \{$\gamma=0, T_{c2}=0.4$\} case for $T>T_{c2}=0.4$, since
then the second band is depleted and only the first band
superconducts.

\subsection{H-T phase diagrams}

\begin{figure}[b]
\includegraphics[width=\linewidth]{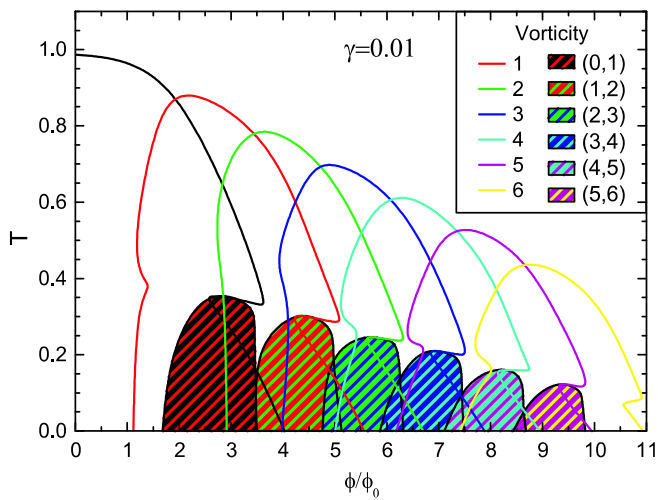}
\caption{\label{HTBoundary} (Color online) The magnetic
flux-temperature stability regions for different vortex states
(vorticity $L$) in a two-band disk of size $R_D/\xi_{10}=4$, and
with $T_{c2}=0.44$, $\delta=1.33$, $\alpha=0.844$ and $\gamma=0.01$.
In color-coded areas, the vortex state is fractional and cannot be
represented by a single $L$, but rather as $(L_1,L_2)$ state, where
vorticities in two condensates are given respectively.}
\end{figure}

In increasing magnetic field, more vortices penetrate the
superconducting system. It is known that the symmetry of the vortex
states is strongly affected by the symmetry of the mesoscopic
sample, as detailed in Refs. \cite{Symmetry}. With increasing
temperature, the symmetry of the sample is even stronger imposed on
the vortex matter, and it is therefore no surprise that in
mesoscopic disks most vortex state configurations collapse into a
giant-vortex at high temperature. We can construct an $H-T$ diagram
for mesoscopic samples, indicating the area of stability of states
with different vorticity. Two-gap systems make there no exception,
but do comprise several particularities. In Fig. \ref{HTBoundary} we
displayed the full stability regions of all possible vortex states
with vorticity $L<7$ in a superconducting disk with parameters
 \{$T_{c2}=0.44$, $\delta=1.33$, $\alpha=0.844$,
$m=1$\},\cite{Chib2Gap} which is very similar to MgB$_2$ except for
the coupling parameter, where we took significantly smaller
$\gamma=0.01$. This choice provides more complexity to the vortex
states, as it allows for different vorticities and vortex
arrangements in the two bands.

\begin{figure}[h]
\includegraphics[width=\linewidth]{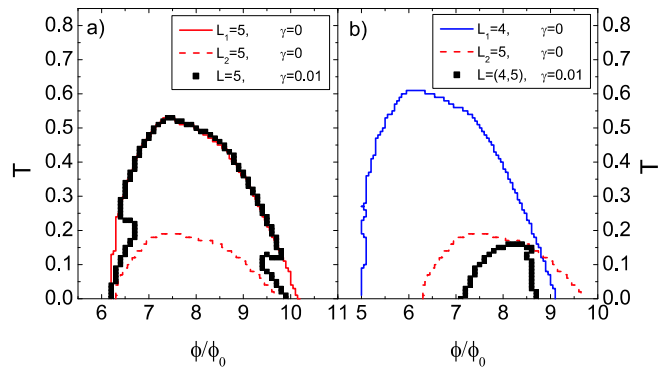}
\caption{\label{HTBound1} (Color online) Illustration on how the
stability flux-temperature regions of integer (a) and fractional
states (b) arise from the single-gap picture. }
\end{figure}
Indeed, one difference from the single-gap superconducting disks is
directly visible in Fig. \ref{HTBoundary}, where the stability
regions of composite vortex states are {\it mushroom shaped}. In
other words, with increasing temperature, one can exit the stability
range of a particular $L$-state, but then find it again at higher
temperatures. This shape has the following origin: At high
temperatures ($T_{c2} \ll T < T_{c1}$) the second gap would be
completely depleted if it wasn't for the coupling. In other words,
the second band depends completely on the first band, and therefore
has the same behavior and features like the H(T)-boundary. However,
for low temperatures ($T \lesssim T_{c2}$) the second band is still
active and retains its own character, and therefore the stability
region boundary follows quite closely the single-gap stability
region.

In the mushroom-shaped areas, the vorticity in the two bands is the
same. However, in the shaded areas we found vortex states where the
vorticity differs from one band to the other. As a consequence, the
overall, apparent vorticity of the sample is no longer integer!
These are the so-called {\it fractional vortex states}.

As clearly shown in Fig. \ref{HTBound1}(a), the full $H-T$ stability
region of an integer flux vortex state is related to the union of
the $H-T$ stability regions for the given state in the two
corresponding single-gap condensates. On the other hand, the
fractional states are found at the intersection of two corresponding
single-gap stability regions [see Fig. \ref{HTBound1}(b), for the
$(L_1,L_2)=(4,5)$ state].

\begin{figure}[h]
\includegraphics[width=0.9\linewidth]{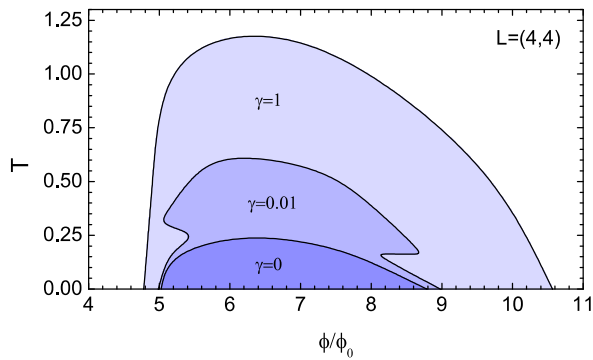}
\caption{\label{HTBoundL44vsGamma} (Color online) The stability
region of the $L=4$ integer flux vortex state, for three different
strengths of coupling $\gamma$.}
\end{figure}

With increasing coupling parameter $\gamma$, the vortex states in
the two condensates are linked together, and moreover reinforce each
other. In Fig. \ref{HTBoundL44vsGamma} we show the stability region
of the integer flux $L=4$ state, for three values of $\gamma$, where
the $H-T$ stability region grows with $\gamma$. We conclude that
increasing $\gamma$ {\it stabilizes} the integer flux states, but
for the same reason {\it destabilizes} the fractional states. We
discuss the latter further in the following section.

\subsection{Fractional vortex states}

\begin{figure}[t]
\includegraphics[width=\linewidth]{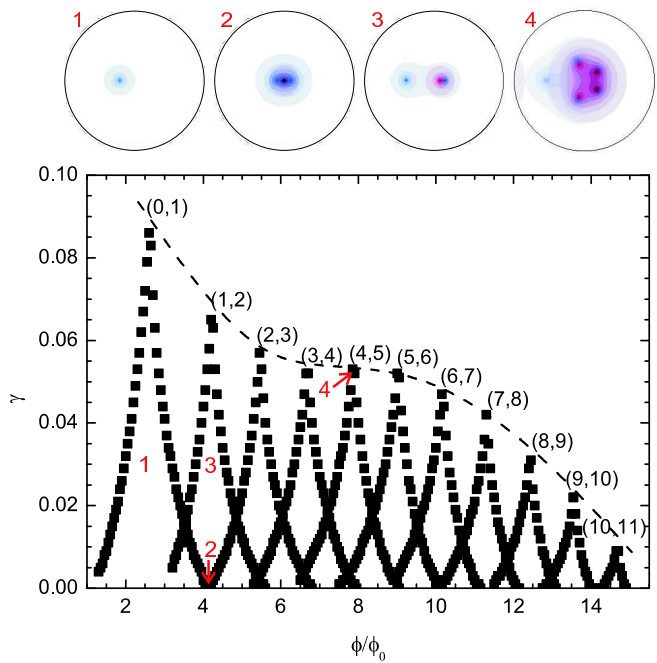}
\caption{\label{PhiGamma} (Color online) The stability regions in
$\phi-\gamma$ parameter space of fractional states with different
vorticities in the two bands. Parameters of the sample are
$R_D/\xi_{10}=4$, $\alpha=\delta=m=1$ and $T=0$. In insets at the
top of the figure, we superimposed the logarithmic plots of the
Cooper pair density in the two gaps on each other (red/blue shades
for condensates 1/2 respectively) for states indicated in the phase
diagram by the red numbers.}
\end{figure}
The existence of fractional states, i.e. states with different
vorticity in the bands, depends strongly on the coupling. They
survive only at weak Josephson coupling between the bands, while
only integer flux states are possible at large $\gamma$ values. This
is illustrated in Fig. \ref{PhiGamma}, where we show that the region
of stability of fractional states shrinks with increasing coupling,
but also that lower vorticity fractional states are more resilient
to $\gamma$. Another interesting aspect of fractional states is
their strong affinity to asymmetry. In both condensates vortices
attempt to form a symmetric shell, but due to coupling and different
respective number of vortices, the final state becomes asymmetric in
most cases. For that reason, the asymmetry is more apparent at
larger coupling $\gamma$. We show several examples through the
$\log$-plots of the Cooper pair density of the chosen states in Fig.
\ref{PhiGamma}. Note that the fractional state not necessarily
contains vortices in both condensates; for example, inset 1 in Fig.
\ref{PhiGamma} is the $(0,1)$ state. Due to coupling, the total
energy is minimized when regions with depleted order parameter in
two condensates are on top of each other. As a result, the vortex of
the second band is attracted to the boundary of the sample, where
the circulating Meissner currents strongly suppress the order
parameter in the first band. Inset 2 is the $(1,2)$ state for
$\gamma=0$, i.e. the condensates are decoupled. This fractional
state is therefore two-fold symmetric, but when we increase $\gamma$
we enhance the asymmetric $(1,2)$ state, as shown in inset 3. One
vortex of the second band is attracted to the vortex of the first
band, and the other is attracted to the edge of the sample. Finally
we show in inset 4 the $(4,5)$ state, at the verge of its stability
region, showing maximally pronounced asymmetry. Four vortices in
both bands sit on top of each other, and the remaining, fifth vortex
of the second band, breaks the symmetry and is gradually pulled out
of the sample. The found states look similar to what was found
earlier for Coulomb bound classical
particles,\cite{CoulombParticles} although underlying physics is
very different.

\begin{figure}[t]
\includegraphics[width=\linewidth]{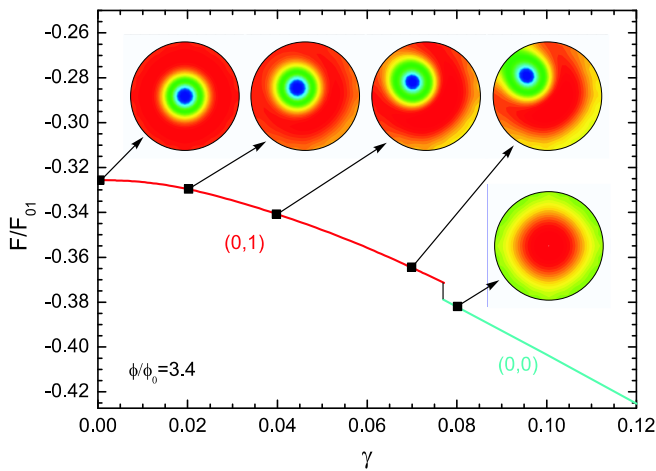}
\caption{\label{EnVSGammaNonComposite01} (Color online) Calculated
free energy of the fractional $(0,1)$ state, as a function of
Josephson coupling between the bands. Insets show contour plots of
the Cooper-pair density in the second gap, illustrating how the
asymmetry gradually increases with $\gamma$ for taken MgB$_2$
parameters and size of the disk $R_D=5\xi_{10}$. The $\gamma$-sweep
was done along the vertical dashed line in Fig. \ref{PhiGammaMgB2}.}
\end{figure}

A two-gap mesoscopic system is a prime example of a vortex system
with competing interactions. Besides the vortex-vortex interactions
in each band, one must take into account the coupling between order
parameters across the bands, and the mesoscopic effect of the
compression of vortices to the interior by the circulating Meissner
current that is maximal at the edge. For example, consider the
$(0,1)$ state, where an outward force originates from the coupling
between the vortex in the second condensate and the suppression of
superconductivity at the edge of the first one. However, this action
competes with the inward force exerted by the Meissner current. This
purely mesoscopic effect leads to a tuneable position of the vortex
in this fractional state: while the Meissner current is roughly the
same at a given magnetic field, the changed coupling between the
condensates brings the vortex further to the boundary. This is shown
in Fig. \ref{EnVSGammaNonComposite01}, as a transition from a
fractional $(0,1)$ vortex state to an integer $L=0$ vortex state
with increasing coupling.

\begin{figure}[bh]
\includegraphics[width=\linewidth]{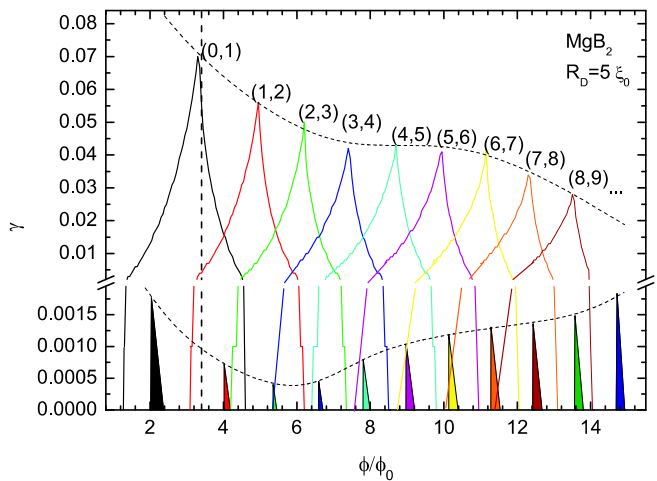}
\caption{\label{PhiGammaMgB2} (Color online) The stability regions
and the ground state (colored) regions in $\phi-\gamma$-space of the
fractional states. Taken parameters are $R_D/\xi_{10}=5$,
$\alpha=0.844$, $\delta=1.33$, $m=1$, $T_{c2}=0.44$ and $T=0.1$,
corresponding to MgB$_2$.}
\end{figure}

As fascinating as they are, the fractional states are difficult to
find in the ground-state. For example, when the coherence lengths of
the two bands are the same, then the energy landscape in both bands
- considered as separate single-gap superconductors - are
proportional, i.e. $F_1 = \alpha\delta^2/m F_2$. All possible vortex
states thus have their ground state in the same phase space region.
The total energy of the system, $F = F_1 + F_2$, will therefore be
proportional to the single-gap energy with as a direct consequence
that fractional states always will have higher (or equal) energy
compared to the integer states. To realize fractional states as the
ground state, one therefore needs to make the discrepancy between
the coherence lengths as large as possible. This can be done by
taking $\alpha$ significantly different from one, or by taking
temperature close to $T_{c2}$, when $T_{c1}>T_{c2}$. In Fig.
\ref{PhiGammaMgB2} we show the stability and ground state regions of
the non-composite states in a disk of size $R=5\xi_{10}$ at
temperature $T=0.4$ (and $\alpha=0.5$).

These asymmetric states can be observed in mesoscopic two-band
samples. As main candidates for such an experiment, we select the
imaging of only the $\pi$-band, as was done recently in Ref.
\cite{PiBandVortexImaging}. Alternatively, scanning Hall
magnetometry or magnetic force microscopy can both reveal the
asymmetric magnetic response of the sample in the case of a
fractional state. We will revisit this point in the section devoted
to magnetic coupling.

\subsection{The superconducting-normal phase boundary}

\begin{figure}[t]
\includegraphics[width=0.8\linewidth]{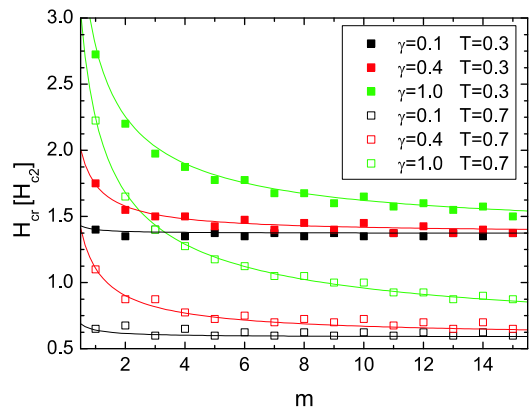}
\caption{\label{mdep} (Color online) (a) The critical field of the
mesoscopic disk $H_{cr}$ as a function of the mass ratio in two
condensates. Parameters used are $R_D=4\xi_{10}$, $T_{c2}=0.5$,
$\alpha=0.1$. (b) {\it Idem} but now as a function of $\alpha$.}
\end{figure}
As mentioned in preceding sections, in Refs. \cite{Drag4,Drag5} an
expression was derived for the second critical field of a bulk two
gap superconductor, given by Eq. (\ref{hc2twogap}). That expression
contains the dependence of the critical field not just on $\gamma$,
but on $m$ and $\alpha$ as well. We have shown that that dependence
does not describe the vortex properties in the two-band samples, but
at this point we check its applicability for the estimation of the
upper critical field of mesoscopic two-band disks at a given
temperature. Our results for the dependence of the upper critical
field on the mass ratio in two bands are shown in Fig.
\ref{mdep}(a), and demonstrate perfect agreement with Eq.
(\ref{hc2twogap}), provided that the found critical field is scaled
by its value at zero temperature and in absence of coupling.

\begin{figure}[h]
\includegraphics[width=\linewidth]{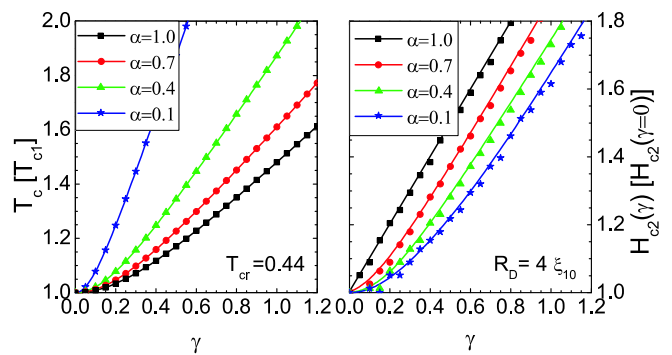}
\caption{\label{TcAndHc2VSGammaZeroFlux} (Color online) (a) The
critical temperature $T_{cr}$ of a two-band mesoscopic disk as a
function of Josephson coupling $\gamma$, in absence of magnetic
field. (b) The upper critical field $H_{cr}$ vs. $\gamma$ for $T =
0$. Dots represent the numerical data, and the solid line is the
result of Eqs. (\ref{hc2twogap}-\ref{Tctwogap}).}
\end{figure}

\begin{figure}[bh]
\includegraphics[width=\linewidth]{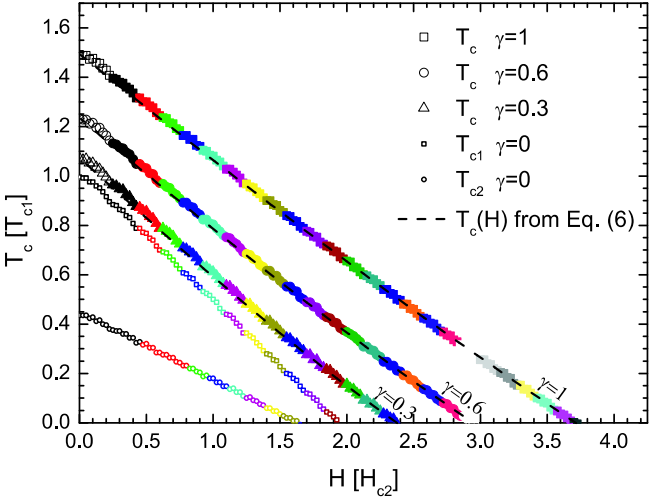}
\caption{\label{TcVSFlux} (Color online) The superconducting-normal
phase $H(T)$ boundary for the sample with parameters $R_D=4\xi_0$,
$\delta=1.33$, $\alpha=0.844$, $T_{c2}=0.44$ and $m=1$, for
different values of the Josephson coupling strength. Different
colors of the dots mean different vorticities. Dashed lines are
obtained from Eq. (\ref{hc2twogap}) with a prefactor described in
the text.}
\end{figure}

Eq. (\ref{hc2twogap}) is further applicable for the estimation of
the $H-T$ superconducting-to-normal phase boundary. Namely, equating
that expression to zero gives the expression for the critical
temperature of the two-band sample:
\begin{equation}
T_{c} = \frac{1}{2} \left( 1 + T_{cr} + \sqrt{(1-T_{cr})^2 +
4\frac{\gamma^2}{m\alpha}T_{cr}}\right). \label{Tctwogap}
\end{equation}
This means that the critical temperature of a two gap superconductor
is always equal or higher than the sum of the critical temperatures
of the two bands, in the case as if there was no coupling. This
observation is in contradiction with findings of Ref.
\cite{Eisterer}, where it is claimed that also a lower $T_c$ is
possible, depending on the parameters. The authors obtained these
results from a microscopic derivation.

Above expressions were originally derived for bulk samples. It is
already known that the upper critical field in mesoscopic
superconductors is higher than in bulk,\cite{CriticalFieldTopology}
and it is therefore intuitively clear that Eq. (\ref{hc2twogap})
would not work for the case of two-gap mesoscopic disks. In Fig.
\ref{TcAndHc2VSGammaZeroFlux} we show the numerically obtained
critical temperature $T_{cr}$ and upper critical field $H_{cr}$
(corresponding to bulk $H_{c2}$) versus $\gamma$ in disks of size
$R=4 \xi_{10}$. We found that both the dependence of critical
temperature and field on $\gamma$ obey the dependencies given in
Eqs. (\ref{hc2twogap}-\ref{Tctwogap}), provided that the critical
field is scaled to its value in the absence of coupling and at zero
temperature. In Fig. \ref{TcVSFlux} we show the calculated
$T_{c}(H)$ boundary for different coupling strengths. Eq.
(\ref{hc2twogap}) can also be inverted to describe the dependence of
$T_{c}$ on the applied field. Although derived for bulk, we find
that latter equation nicely fits the $H(T)$-curves in Fig.
\ref{TcVSFlux} for a mesoscopic disk, after the aforementioned
scaling of the magnetic field.

\section{Magnetic coupling}

In the previous section, we assumed the existence of a Josephson
coupling between two superconducting bands, but we neglected the
screening of the magnetic field. In applied magnetic field, the
magnetic response of a two-band superconductor follows from the
induced supercurrent:
\begin{eqnarray}
-\kappa_{1}^{2} \Delta \vec{A} = \vec{j}_s & = & \Re\left[
\psi_1\left( i\nabla - \vec{A}\right)\psi_{1}^{*} \right] \nonumber \\
& + & \frac{m}{\delta^2} \Re\left[ \psi_2\left( i\nabla -
\vec{A}\right)\psi_{2}^{*} \right].
\end{eqnarray}
Conventionally, the demagnetization and screening effects in
mesoscopic superconductors are expressed through the Ginzburg-Landau
parameter $\kappa$, being equal to the ratio of penetration depth
$\lambda$ and coherence length $\xi$. For that reason, we
reformulate the equations of section (\ref{glequat}) to introduce
$\kappa_2$, the GL parameter of the second condensate instead of the
parameter $\delta$, the ratio of the order parameters in two bands.
We start from the definitions of $\xi$ and $\lambda$:
\begin{eqnarray*} \lambda_{n0}^{2} = \frac{m_n c^2 \beta_n}{16 \pi
\alpha_{n0}e^2} & \hspace{20pt}& \xi_{n0}^{2} = \frac{\hbar^2}{2 m_n
\alpha_{n0}},
\end{eqnarray*}
to derive:
\begin{equation}
\label{k1k2malphadelta} \frac{\kappa_{1}^{2}}{\kappa_{2}^{2}} =
\frac{m}{\delta^2 \alpha},
\end{equation}
which we then substitute in the GL equations to obtain:
\begin{eqnarray}
(-i\nabla - \vec{A})^2 \psi_1 &-& (1 - T - |\psi_1|^2) \psi_1 =
\frac{\gamma\alpha}{m}\frac{\kappa_1}{\kappa_2} \psi_2, \label{GL1}\\
(-i\nabla - \vec{A})^2 \psi_2 &-& \alpha \left(1-\frac{T}{T_{c2}} -
|\psi_2|^2 \right) \psi_2  =
\frac{\gamma}{\sqrt{m\alpha}}\frac{\kappa_2}{\kappa_1}\psi_1,\label{GL2}
\\
-\Delta \vec{A} = \vec{j}_s &=& \frac{1}{\kappa_{1}^{2}} \Re\left[
\psi_1\left( i\nabla - \vec{A}\right)\psi_{1}^{*} \right] \nonumber \\
&+&\frac{\alpha}{\kappa_{2}^{2}} \Re\left[ \psi_2\left( i\nabla -
\vec{A}\right)\psi_{2}^{*} \right].\label{VecPot}
\end{eqnarray}

\begin{figure}[h]
\includegraphics[width=\linewidth]{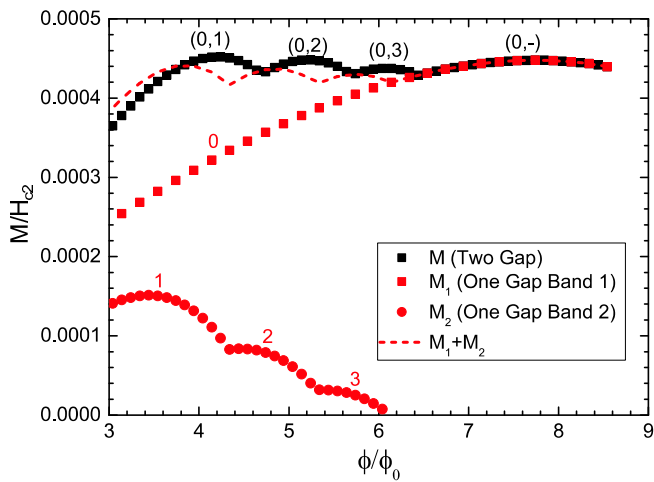}
\caption{\label{Mag2vs1Gap} (Color online) Magnetization versus
applied magnetic field ($M(H)$) loops for a two-gap disk with radius
$R_D=10\xi_{10}$, and two condensates as single-gap samples, in the
absence of Josephson coupling. Magnetic coupling is included in the
calculations, with parameters $\kappa_1=3.68$, $\kappa_2=0.66$, and
$\alpha=0.06$.}
\end{figure}

This form of two-band Ginzburg-Landau equations is particularly
convenient for comparison with the conventional types of
superconductivity. In the single-gap bulk samples, the value of
$\kappa$ above or below $1/\sqrt{2}$ determines the superconductor
being of second or first type, respectively. For a two-band sample,
this distinction is much more difficult to establish, since Eqs.
(\ref{GL1}-\ref{VecPot}) show the direct influence of not only
$\kappa_1$ and $\kappa_2$, but also the Josephson coupling and the
squared ratio of coherence lengths in two condensates $\alpha$.

Eq. (\ref{VecPot}) also shows that two bands are directly coupled
through the screening currents, and this type of coupling we refer
to as magnetic coupling. In Fig. \ref{Mag2vs1Gap} we show the
calculated magnetization of the disk with radius $R=10\xi_{10}$ as a
function of the applied field (in absence of Josephson coupling), to
illustrate how the magnetic field couples the two bands. We observe:
(i) the magnetization of the coupled system is somewhat higher than
the sum of the two uncoupled systems; (ii) the found cusps are
wider, since the Meissner currents in one band screens the field in
the other band and flux entry is therefore more difficult. The net
field in the interior of the sample is therefore lower, and the
critical field is increased. This is in accordance with existing
experimental findings on MgB$_2$.

However, the information available in literature is also often
confusing. For example, we identified two sets of parameters, both
believed to be correct for MgB$_2$. From Refs. \cite{Mosh,
TypeIXSurfaceEnergy} we extracted $\kappa_1=3.68$, $\kappa_2=0.66$,
$\alpha=0.068$ which should be valid for a clean sample (single
crystal). In these works the strength of the Josephson coupling,
$\gamma$, is not estimated. In the dirty limit, the compound is
definitely a type-II material. Substituting former values into Eq.
(\ref{k1k2malphadelta}) we obtain $\delta^2/m \approx 0.59$. On the
other hand, from Refs. \cite{Chib2Gap,Drag5} we obtain $T_{c2}=0.56
T_{c1}$, and $\delta=1.33$. For usual Mg$^{11}$B$_2$ we have
$\gamma=0.4$, $m=1$ and $\alpha=0.844$. For the Mg$^{10}$B$_2$ we
found $\gamma=0.28$. For irradiated MgB$_2$ samples a mass ratio
$m\approx14$ has been observed together with
$\alpha=0.059$.\cite{Chib2Gap} Therefore, in the remainder of the
paper we will not restrict ourselves just to the particular values
of the parameters, but rather focus on new physics between the two
types of superconductivity and its manifestations.

\subsection{Magnetic vs. Josephson coupling}
\begin{figure}[t]
\includegraphics[width=\linewidth]{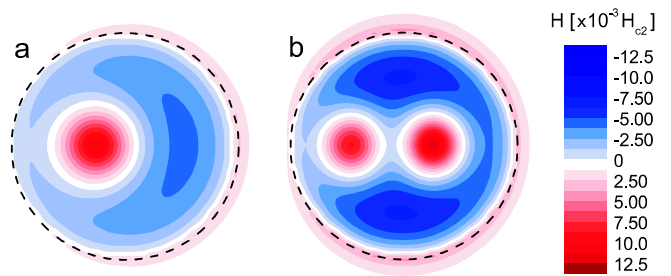}
\caption{\label{NonCompositeMagField} (Color online) The magnetic
response of the fractional (0,1) and (1,2) vortex states in a
mesoscopic two gap superconducting disk with parameters $R_D=10
\xi_{0}$, $\kappa_1=10$, $\kappa_2=2$, $\alpha=0.3$, $\gamma=0.01$
and $T=0$. The applied field in (a) is $0.04 H_{c2}$
($\phi/\phi_0=2$) and in (b) $0.08 H_{c2}$ ($\phi/\phi_0=4$). The
dashed line shows the sample boundary. }
\end{figure}

\begin{figure}[t!]
\includegraphics[width=0.88\linewidth]{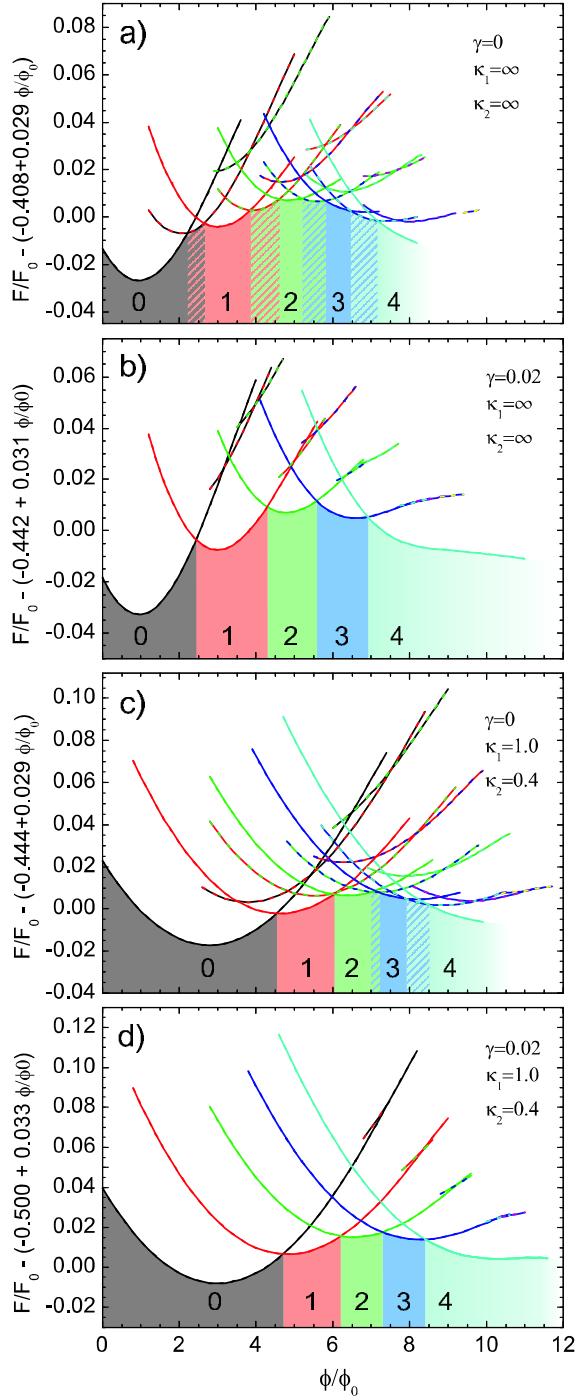}
\caption{\label{FreeEnNonComposite1} (Color online) Free energy as a
function of magnetic flux for found integer and fractional states
with $L \leq 4$. A linear background is subtracted from all curves
to enhance readability (as indicated in the labels). The colors of
the curves correspond to the vorticity, and also indicate the
combinations involved in the (two-colored) fractional states. The
ground state is indicated by the color-coded shaded areas below the
curves. Parameters of the sample are $R_D = 7.5 \xi_0$, $T=0.2$,
$T_{cr}=0.5$, $\alpha=0.3$, and $m=1$.}
\end{figure}

\begin{figure}[t]
\includegraphics[width=\linewidth]{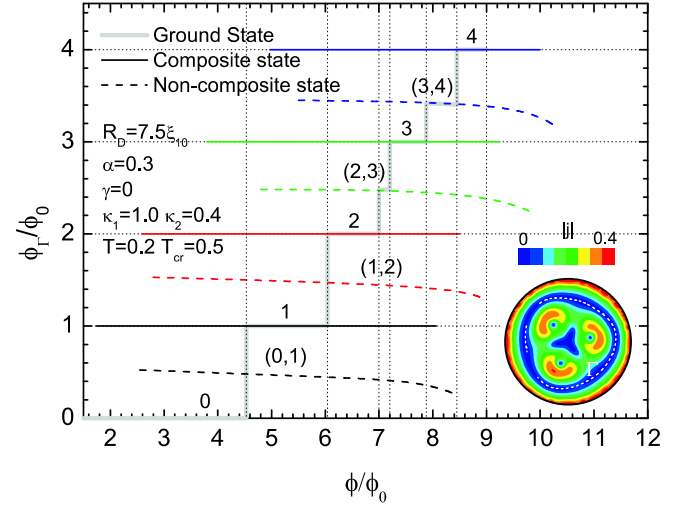}
\caption{\label{FluxNoCoupling} (Color online) The flux quantization
in a two gap superconductor bands only coupled by the magnetic
field. $\phi_{\Gamma}$ is the flux measured through the contour
$\Gamma$, defined as the contour where the supercurrent equals zero,
as illustrated by the white dashed line in the inset. The thick
solid grey line in the graph depicts the ground-state as a function
of applied field.}
\end{figure}

As mentioned above, one of the most fascinating properties of a
two-band system is the possible appearance of fractional and
fractional states. We argued that those could be observed in
experiment through their magnetic response. Using the preceding
theoretical formalism, we can now calculate the magnetic field in
and around the sample, in response to the applied magnetic field. In
Fig. \ref{NonCompositeMagField} we show the magnetic field profile
emanating from the mesoscopic superconducting disk in the fractional
(0,1) and (1,2) vortex state. The asymmetry induced by the Josephson
coupling is clearly visible and can be directly imaged in magnetic
force microscopy or scanning Hall probe magnetometry experiments.
Additionally, the integrated magnetic field from such measurements
will reveal the fractional flux carried by these states of
non-integer total vorticity.

In Fig. \ref{FreeEnNonComposite1} we show the effect of both
Josephson and magnetic coupling on the free energy and the vortex
states. Fig. \ref{FreeEnNonComposite1}(a) shows the energy landscape
when superconducting condensates are decoupled. As could be
expected, all $(L,L+1)$ fractional states are stabilized in the
ground state, at intermediate fields between integer flux $L$ and
$L+1$ states. $(L,L+2)$ states also exist, but have significantly
higher energy. The same holds for $(L+1,L)$ fractional states,
having far higher energy than the corresponding $(L,L+1)$ state.
This can be entirely inverted for a different choice of parameters,
particularly $\alpha$, which determines the relative coherence
lengths and consequently the ratio of the vortex energy in the two
bands. In the present calculation, we therefore omit the curves
corresponding to $(L+1,L)$ states.

In Fig. \ref{FreeEnNonComposite1}(b) the Josephson coupling is
added. This directly results in stabilization of the integer flux
states in the ground state, and fractional ones have much higher
energy. Generally, the $(L,L+n)$ energy increases further and energy
levels follow each other as magnetic field and $n$ are increased.

In Fig. \ref{FreeEnNonComposite1}(c) we introduced the magnetic
coupling, in absence of the Josephson one. This broadens the
stability intervals of all vortex states - integer and fractional -
as a consequence of the magnetic screening which lowers the
effectively experienced field by the sample and enhances
superconductivity. However, the magnetic response of the sample is
generally of oblate shape (due to the symmetry of the disk), and
asymmetric states are less favorable than in case (a). For that
reason, pronouncedly asymmetric (0,1) and (1,2) states are not
present in the ground state. However, higher fractional states can
be found in the ground state since the ring-like arrangement of
vortices and their fields in both condensates enhance each other,
and also follow better the overall symmetry of the stray magnetic
field. Nevertheless, Josephson coupling is still able to completely
remove the fractional states from the ground state, as shown in Fig.
\ref{FreeEnNonComposite1}(d). Arguably, at higher vorticity
fractional states could appear in the ground state, since the order
parameters in the two bands will become increasingly similar with
increasing vorticity.

In Fig. \ref{FluxNoCoupling} we illustrate the flux-quantization in
a two gap superconducting disk. As is already known, the flux in
single-band mesoscopic samples is not quantized, but it always is
within a contour determined by the zero current. Therefore we
compute the flux $\phi_{\Gamma}$ penetrating the two-band sample
through a contour $\Gamma$ inside the sample on which $\vec{j}_s=0$.
The result is plotted in the presence of magnetic coupling between
the bands to still have some fractional vortex states left in the
ground state. By plotting the whole stability regions of the states,
we noticed that the fractional flux decreases with increasing
applied magnetic field (i.e. the applied flux $\phi$). We pinpoint
this effect in addition to the findings of Ref. \cite{Chib2Gap},
where the authors found the decrease of flux through contour
$\gamma$ with increasing temperature. The reason for the change of
the fractional flux with applied field or temperature is that one of
the condensates always depletes faster than the other (in the
present case, the second one). At sufficiently high field or
temperature, only one band superconducts, and the flux through
contour $\Gamma$ changes towards its quantized level in the
surviving band. In the present case, the first band is stronger, and
the fractional flux decreases towards vorticity in the first band,
i.e. $L=1$. If the considered state was a $(2,1)$ one, the
fractional flux would increase towards $L=2$ level with increasing
field or temperature.

\subsection{Magnetization curves}

\begin{figure}[b]
\includegraphics[width=\linewidth]{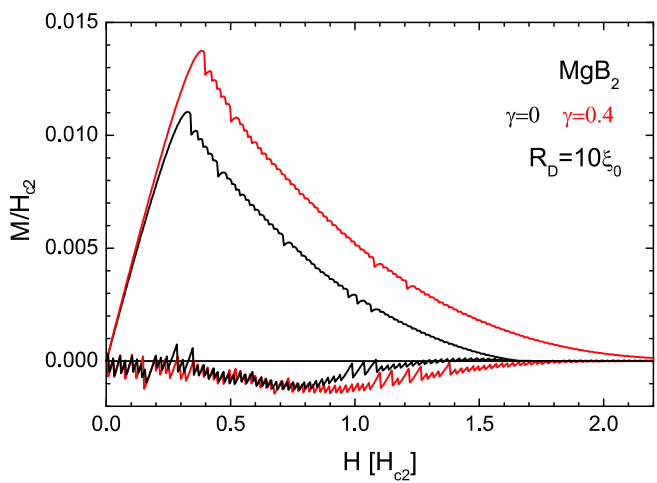}
\caption{\label{MHLoopMgB2.eps} (Color online) $M(H)$ loops obtained
by sweeping up and down the magnetic field for a MgB$_2$
single-crystal disk of radius $R_D=10\xi_{10}$.}
\end{figure}
We define the magnetization as: $M = \frac{1}{4\pi V} \int \Delta
H_z dV$, where $\Delta H$ represents the induced magnetic field. In
Fig. \ref{MHLoopMgB2.eps} we plot the $M(H)$-loop in a MgB$_2$ disk
with radius $R_D=10\xi_0$ and thickness $d=\xi$ at $T=0$, taking the
parameters of Ref. \cite{Mosh}, i.e. $\kappa_1=3.68$,
$\kappa_2=0.66$, $\alpha=0.068$. Although the second band is
conventionally type-I, the shape of the $M(H)$-loop suggests that
the whole system still behaves like a type-II superconductor, i.e.
there is no indication of a type 1.5 superconductivity reported by
Moshchalkov {\it et al}.\cite{Mosh} We additionally plotted the
magnetization in the absence of Josephson coupling, which also does
not show any qualitative deviation of type-II behavior. The only
influence of the Josephson coupling is an apparent increase of
critical field and a stronger magnetic response, which is a direct
consequence of currents being strengthened by coupling.
\begin{figure}[b]
\includegraphics[width=\linewidth]{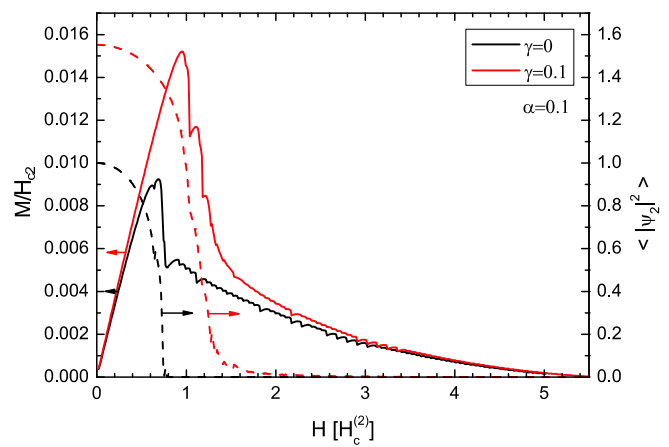}
\caption{\label{MHLoopsGammaDuo} (Color online) Solid curves
represent magnetization as a function of applied magnetic field for
$\gamma=0$ (black) and $\gamma=0.1$ (red), and parameters of the
sample $\kappa_1=1$, $\kappa_2=0.2$, $\alpha=0.1$, $R_D = 10
\xi_{10}$ and $T=0$. Dashed curves represent the mean Cooper pair
density in the second condensate. }
\end{figure}
However, when we take lower values of the GL parameters, for example
$\kappa_1=1$ and $\kappa_2=0.2$, we find, as shown in Fig.
\ref{MHLoopsGammaDuo} a behavior of the magnetization versus field
that is neither type-II nor type-I like. This state is characterized
by a steep drop of the magnetization at a field close to the
thermodynamical critical field $H_c$ of the second condensate
($H_{c}^{(2)}$). Due to finite demagnetization effects,
characteristic of type-I samples, the transition is at a lower field
than $H_{c}^{(2)}$. At this transition field, superconductivity
ceases in the second gap, and the magnetization undergoes a steep
drop. The origin of this effect is clearly visible in the figure,
where also the mean Cooper pair density in the sample is plotted -
the magnetization drop coincides with the depletion of the second
band. Beyond the transition field, the flux continues to enter the
sample gradually, exhibiting the type-II mixed state of the first
condensate, and the overall behavior of magnetization can be treated
as a superposition of type-I (steep drop) and type-II (gradual
decrease) behavior of the two condensates, each being of different
type. The influence of the Josephson coupling is also striking, as
it smoothes out the drop in the magnetization: the second condensate
still depletes but at a slower rate due to the exchange of Cooper
pairs with the first condensate. The slope of the decrease of the
mean Cooper pair density in the second condensate still seems to
match the slope of the drop in magnetization with a remarkable
accuracy, although the transition becomes less abrupt and more
reminiscent of a type-I intermediate state with bundles of flux
penetrating the sample.
\begin{figure}[t]
\includegraphics[width=\linewidth]{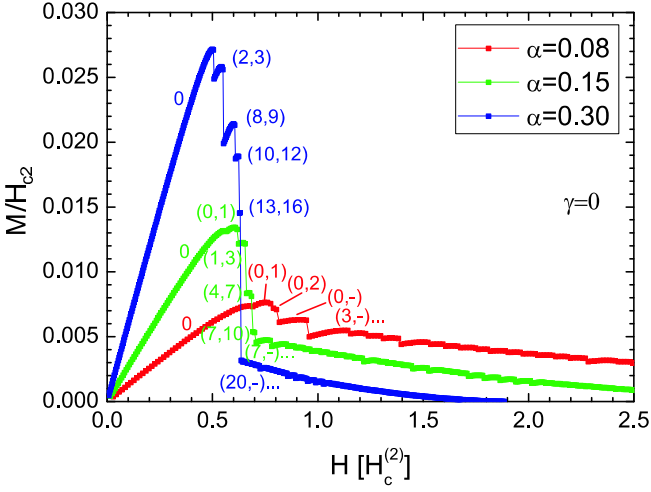}
\caption{\label{MHGam0MultiAlfa} (Color online) Magnetization of the
sample versus the applied field for different values of the ratio
between the coherence lengths in two condensates $\alpha$. Other
parameters of the sample are $\gamma=0$ $\kappa_1=1$,
$\kappa_2=0.2$, $R_D = 10 \xi_{10}$ and $T=0$.}
\end{figure}
\begin{figure}[t]
\includegraphics[width=\linewidth]{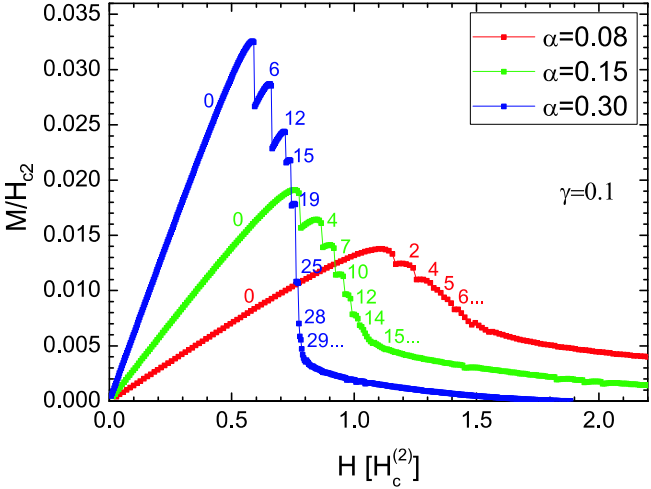}
\caption{\label{MHGam0.1MultiAlfa} (Color online) {\it Idem} as Fig.
\ref{MHGam0MultiAlfa}, but for finite Josephson coupling
$\gamma=0.1$. }
\end{figure}

In Fig. \ref{MHGam0MultiAlfa} and \ref{MHGam0.1MultiAlfa} we
demonstrate the influence of the ratio of the coherence lengths,
$\alpha$ on the magnetic behavior of the sample. In Eq.
(\ref{VecPot}), the supercurrents due to the second condensate have
a prefactor of $\alpha/\kappa_{2}^{2}$. Therefore, it is not
$\kappa_2$ alone which determines the type of the band and an
effective $\kappa_{2}^{eff}=\kappa_{2}/\sqrt{\alpha}$ can be
introduced. Essentially, this suggests that the self induced field
should be proportional to $\alpha$, and the results in Fig.
\ref{MHGam0MultiAlfa} are supportive of this. The level of
magnetization to which the sample jumps after the magnetization drop
should be independent of $\alpha$, since the second condensate is
depleted there. However it turns out that the larger the
magnetization is before the drop, the lower it becomes after the
drop. This follows from the fact that, before the magnetization
drop, the second condensate is able to provide a better screening of
the magnetic field for the first condensate when $\alpha$ is larger,
but when the second condensate ceases, the first condensate
experiences a large difference in the felt magnetic field, which in
turn allows for a larger flux penetration and thus a lower
diamagnetic response of the sample.

It should be stressed here that the field at which the second
condensate depletes and the magnetization drops is also influenced
by $\alpha$. The smaller the coherence length of the second
condensate (higher $\alpha$), the smaller the transition field. This
difference is even more prominent in the presence of Josephson
coupling. In other words, the apparent demagnetization effect in the
type-I part of the magnetization curve is clearly influenced by the
parameter $\alpha$.

\begin{figure}[b]
\includegraphics[width=\linewidth]{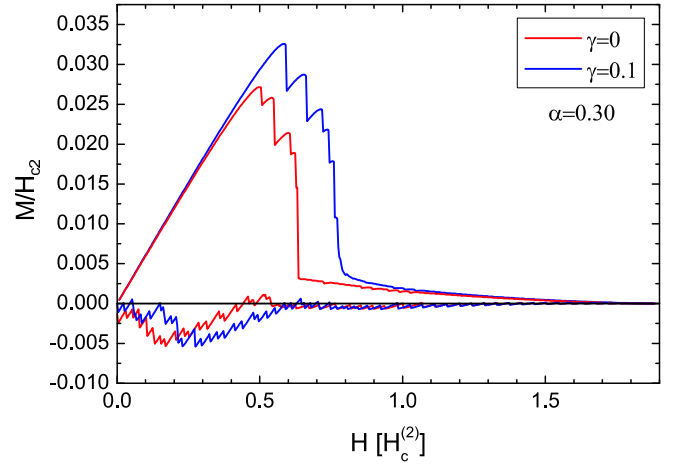}
\caption{\label{MHLoopsAlfa0.3DuoGamma} (Color online) Full sweep up
and down $M(H)$ loops with and without coupling. Taken parameters
are $\alpha=0.3$ $\kappa_1=1$, $\kappa_2=0.2$, $R_D = 10 \xi_{10}$
and $T=0$.}
\end{figure}

Finally, in Fig. \ref{MHLoopsAlfa0.3DuoGamma} we also show the
magnetization corresponding to the sweep-down of the applied
magnetic field. At the point where the second condensate revives the
magnetization jumps up, since there the type-I condensate
contributes to the diamagnetic signal. This jump is less abrupt when
$\gamma$ is non-zero.

\subsection{Type-I.x vortex states}

\begin{figure}[th]
\includegraphics[width=\linewidth]{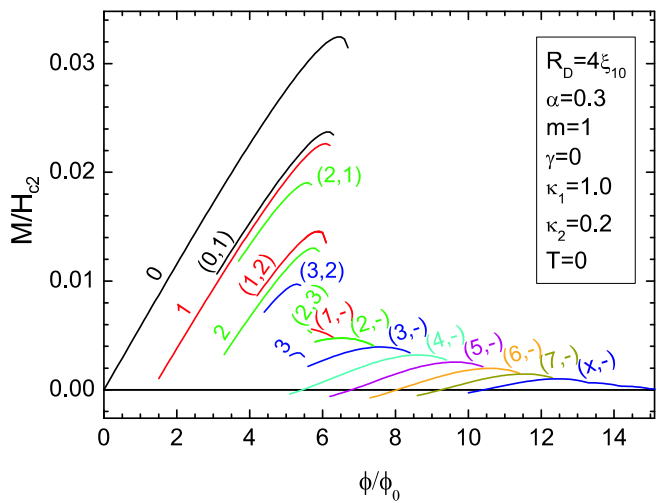}
\caption{\label{RealMHLoops} (Color online) $M(H)$ for all stable
vortex states in a mesoscopic type-I.x superconducting disk.}
\end{figure}

\begin{figure}[t]
\includegraphics[width=0.75\linewidth]{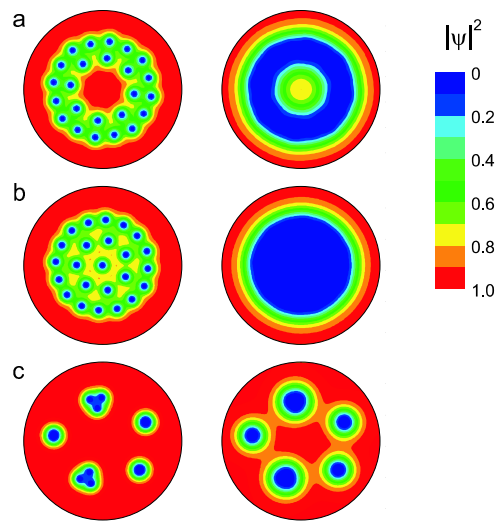}
\caption{\label{vortexstates} (Color online) Cooper pair density
contour plots of the first (left) and second (right) condensate.
$\kappa_1=1$, $\kappa_2=0.2$, $\alpha=0.3$, $R_D=15 \xi_{10}$. In
(a) and (b) $\phi/\phi_0=30$, $L=24$, $\gamma=0$. In (c)
$\phi/\phi_0=15$, $L=12$, $\gamma=0.02$. }
\end{figure}

It is already known that vortices repel each other in type-II
superconductors, form Abrikosov lattice in bulk samples, but are
compressed into geometry dependent multi-vortex states and even
giant vortex states in mesoscopic superconductors. In type-I samples
however, flux penetrates the sample in the form of lamellae or
tubular flux bundles.\cite{Golib} As we showed above, two band
superconductors can show a bit of both behaviors, called type-1.5
superconductivity by Moshchalkov {\it et al.}.\cite{Mosh} While we
demonstrated the type I.x behavior through the magnetization loops,
they mainly discussed the vortex-vortex interaction in two-band
superconductors, claiming that it should be short range repulsive
and long range attractive. This of course assumes integer flux
vortex states, or strong Josephson coupling in our model. However,
as we have seen above, a plethora of other vortex states are
possible, not all with integer flux. Therefore, the vortex-vortex
interaction should be discussed separately within bands
(intra-band), and separately between them (inter-band). While
leaving the detailed analysis for the future, we here show several
prime examples of vortex states that can be found in two-band
mesoscopic disks, that show type-I.x behavior (however different
from Ref. \cite{Mosh}). For example, we take the disk with radius
$10\xi_{10}$ with parameters of the condensates $\kappa_1=1$,
$\kappa_2=0.2$ and $\alpha=0.3$. As shown in Fig.
\ref{vortexstates}(a-b), at larger applied field providing 30 flux
quanta through the sample, we found a $L=24$ vortex state. Due to
absence of Josephson coupling, two allotropic modifications of the
vortex state were found possible. In both, the vortices in type-I
condensate merge into a single domain, either a ring domain (a), or
a giant-bubble (b), which is typical for type-I samples. Actually
they are still single vortices, but have huge overlap, mimicking a
normal domain. Vortices in the type-II condensate remain separate
however, but are forced by magnetic coupling to obey the symmetry of
the intermediate state of the type-I condensate. They therefore form
more or less a conventional multi-shell state under the giant vortex
in (b), but are forced to make an unconventional state with dense
shells within the ring domain in (a). This type-I-II competition is
even more pronounced in the presence of Josephson coupling, as shown
in Fig. \ref{vortexstates}(c). In the first condensate, one can see
three giant $L=2$ vortices and two $L=3$ multi-vortex clusters. This
is also the situation in the second condensate, however the
multi-vortices are now even closer and also overlap more, mimicking
perfect giant vortices. Due to the Josephson coupling, both
condensates influence each other; as a consequence, the type-I
intermediate state is forced to split into as many bubbles as
possible, and vortices in type-II condensate must group into those
bubbles. As a result, a multi-vortex of multi-vortices is formed,
clearly a signature of type-I.x behavior.

\section{Conclusions}
In summary, we presented a theoretical Ginzburg-Landau (GL) study of
the superconducting state of two-band mesoscopic disks, where both
the influence of Josephson and of magnetic coupling between the
superconducting bands are discussed. In cases when screening of the
magnetic field can be neglected, we found the dependence of the size
of the vortex core on the strength of the Josephson coupling and
showed that it generally has an influence opposite to the one of
temperature. In limiting cases, our numerical findings agree well
with analytic expressions available in literature. We also found a
fitting function, which gives an excellent estimate of the size of
the vortex core as a function of the size of the mesoscopic disk. In
our further analysis of the vortex states, we focussed mainly on
exotic, fractional states, where two condensates comprise different
number of vortices and the apparent total vorticity of the sample is
fractional. We reported asymmetric vortex states following from
competing interactions in the two-band mesoscopic system, and showed
how some states can be manipulated by e.g. coupling between the
bands. We indicate how such states can be experimentally observed.
Fractional states can even be found in the ground state, but
typically far from the S/N boundary. We give the expression for the
upper critical field of a two-band mesoscopic disk as a function of
temperature, which is similar to analytic estimations for bulk,
however scaled to its value at zero temperature for zero coupling
between the condensates.

When magnetic screening and coupling between the bands is included
in the simulations, we characterized the response of the sample
through the competition of the GL parameters of the two-bands (with
special attention to the case when one band is type-II and the other
is type-I). However, we show that this is insufficient, and that
Josephson coupling and the ratio of the coherence lengths in the two
bands also play an important role. Although we did not find evidence
for type-1.5 superconductivity in clean MgB$_2$ disks, we did find
its manifestation for a different choice of relevant parameters. The
magnetization vs. applied field shows a distinct jump at the field
where type-I condensate ceases, and the overall shape of the curve
can surely be characterized as type-I.x like. This is also evident
in the found vortex states in the latter case, which are a
combination of single vortices and lamellar domains.

\section*{Acknowledgements}
This work was supported by the Flemish Science Foundation
(FWO-Vlaanderen), the Belgian Science Policy (IAP), the ESF
`Nanoscience and Engineering in Superconductivity' (NES) program,
and the ESF `Arrays of Quantum Dots and Josephson Junctions'
network.

\end{document}